\documentclass[twoside,11pt]{article}
\usepackage{blindtext}
\usepackage{jmlr2e}
\newcommand{\iid}{\stackrel{\mathrm{iid}}{\sim}}
\newcommand{\ind}{\stackrel{\mathrm{ind}}{\sim}}

\usepackage[margin=1.25in]{geometry}

\RequirePackage[OT1]{fontenc}
\usepackage{amssymb, amsmath, natbib, mathabx}
\usepackage{bm}
\usepackage{graphicx}
\usepackage{float}
\usepackage{subcaption}
\usepackage{multicol}
\usepackage{xr}
\usepackage{mathtools}
\usepackage{comment}
\usepackage{nameref}
\usepackage{natbib}
\usepackage{setspace}
\usepackage{verbatim}
\usepackage{placeins} 
\usepackage{xcolor}

\newcommand{\Var}{\mathrm{Var}}

\usepackage{lastpage}
\jmlrheading{}{2024}{1-\pageref{LastPage}}{7/24; }{}{}{Terrance D. Savitsky and Julie Gershunskaya}

\ShortHeadings{Simulation-based Calibration of Uncertainty Intervals}{Savitsky and Gershunskaya}
\firstpageno{1}

\begin{document}

\title{Simulation-based Calibration of Uncertainty Intervals under Approximate Bayesian Estimation }

\author{\name Terrance D.\ Savitsky \email Savitsky.Terrance@bls.gov \\
       \addr Office of Survey Methods Research\\
       U.S. Bureau of Labor Statistics\\
       2 Massachusetts Ave NE\\
       Washington, DC 20212, USA
       \AND
       \name Julie\ Gershunskaya 
       \email Gershunksaya.Julie@bls.gov \\
       \addr Office of Employment and Unemployment Statistics\\
       U.S. Bureau of Labor Statistics\\
       2 Massachusetts Ave NE\\
       Washington, DC 20212, USA}

\editor{}

\maketitle

\begin{abstract}
The mean field variational Bayes (VB) algorithm implemented in Stan is relatively fast and efficient, making it feasible to produce model-estimated official statistics on a rapid timeline. Yet, while consistent point estimates of parameters are achieved for continuous data models, the mean field approximation often produces inaccurate uncertainty quantification to the extent that parameters are correlated a posteriori.  In this paper, we propose a simulation procedure that calibrates uncertainty intervals for model parameters estimated under approximate algorithms to achieve nominal coverages.    Our procedure detects and corrects biased estimation of both first and second moments of approximate marginal posterior distributions induced by any estimation algorithm that produces consistent first moments under specification of the correct model.    The method generates replicate datasets using parameters estimated in an initial model run.   The model is subsequently re-estimated on each replicate dataset, and we use the empirical distribution over the re-samples to formulate calibrated confidence intervals of parameter estimates of the initial model run that are guaranteed to asymptotically achieve nominal coverage.   We demonstrate the performance of our procedure in Monte Carlo simulation study and apply it to real data from the Current Employment Statistics survey.  
\end{abstract}


\section{Introduction}

The Bureau of Labor Statistics (BLS) publishes employment estimates for geography-by-industry domains on a monthly basis in the Current Employment Statistics (CES) survey.  Both a point and associated variance estimate is constructed for each domain from underlying survey data (linked to that domain) acquired from business establishments.  These survey-based domain-level point and variance estimates contain both sampling and measurement sources of error.   It is common practice to conduct small domain estimation modeling using a Bayesian hierarchical formulation \citep{rS1} that borrows information among correlated domains and outputs smoothed domain point and variance estimates.   The Bayesian hierarchical model used by BLS to produce CES domain estimates are richly parameterized (with the number of parameters on the order of the number of domains) to capture the correlation structures; for example, the model parameterizes domain-level random effects under a correlated prior distribution  to capture dependence in the response variable of interest among the domains.   

As a result of the rich parameterization, the resulting computation time for an Markov Chain Monte Carlo (MCMC) estimation algorithm can be too long to support a monthly production process.  Stan \citep{stan:2015} provides an implementation of variational Bayes (VB) in their Automatic Differentiation Variational Inference (ADVI) algorithm that uses optimization instead of the numerical integration of MCMC such that the computation scales better to larger datasets.  BLS utilize Stan's ADVI to produce the CES monthly smoothed estimators from a Bayesian hierarchical model (that we review in the sequel).   

The ADVI algorithm produces very accurate smoothed point estimates for parameters under the BLS model for the CES \citep{r3}, but the resulting variances or scales of the estimated marginal posterior distributions are inaccurate.   BLS desires to publish both the smoothed point estimate, $\mathbb{E}\left(\theta_{i}\mid \mathbf{y}\right)$ of the posterior distribution for domain, $i \in (1,\ldots,N)$ for dataset, $\mathbf{y} = (y_{1},\ldots,y_{N})$, its associated variance, $V(\theta_{i}\mid \mathbf{y})$, along with associated uncertainty intervals for any level of misspecification $\gamma$.  The ADVI estimation algorithm often produces inaccurate scale estimates of the posterior distributions for $\{\theta_{i}\}_{i=1}^{N}$ because it uses a mean-field algorithm that assumes the model parameters are \emph{a posteriori} independent, which is typically not the case.

We introduce a simulation-based procedure in this paper that corrects or calibrates the scale of the marginal posterior parameter distributions and outputs confidence intervals that achieve nominal coverage under any misspecification $\gamma$. Our simulation-based calibration method may be used to calibrate uncertainty quantification under any estimation algorithm for the marginal posterior distributions under two conditions: 1.  The specified estimation model is asymptotically correct (for data generation); 2. The estimation algorithm produces consistent point estimates, though those point estimates may asymptotically express a constant bias. For example, if the algorithm tends to produce an overly small posterior variance for $\theta_{i}$ under an asymptotically correct model specification, then if we generate a replicate dataset indexed by $\alpha$, $\mathbf{y}^{(\alpha)}$,  using posterior draw $\theta_{i}^{(\alpha)}$ as ``truth" and re-estimate the model on $\mathbf{y}^{(\alpha)}$, then the estimated posterior variance of $\theta^{(\alpha)}_{i}$ from this model will also be downwardly biased. The estimation error propagates from the real data to replicate data because the error is induced by the estimation algorithm.

Let index $i \in (1,\ldots,N)$ index a domain. Our simulation-based calibration procedure takes a draw, $\alpha \in (1,\ldots,A)$, of $(\theta^{(\alpha)}_{i})_{i=1}^{N}$ and generates a replicate dataset, $\mathbf{y}^{(\alpha)} = \left(y^{(\alpha)}_{1},\ldots,y^{(\alpha)}_{n}\right)$.  We re-estimate the model on each $\mathbf{y}^{(\alpha)}$ from which we compute the approximate posterior means and variances, $(m(\theta^{(\alpha)}_{i}),v(\theta^{(\alpha)}_{i}))$ from $\mathbf{y}^{(\alpha)}$.   We construct a pivot statistic indexed for each re-sample, $\alpha$, from $(m(\theta^{(\alpha)}_{i}),v(\theta^{(\alpha)}_{i}))$ by treating the point estimate, $\theta^{(\alpha)}_{i}$, as ``truth''.  We use the distribution of this pivot statistic to over the $\alpha \in (1,\ldots,A)$ replications to construct a mean (shift) and variance (scale) adjustment. Each adjusted mean and variance for domain $i$ are subsequently used to formulate a calibrated confidence interval for $\theta_{i}$ that replaces the approximate posterior distributions from the ADVI algorithm.  We enumerate the details of our simulation-based calibration procedure in the sequel.

\subsection{Literature Review}

Our calibration procedure directly operates on the marginal posterior distributions for model parameters and outputs corrected variances and and associated confidence intervals that are asymptotically guaranteed to achieve nominal coverage.

By contrast, conformal inference \citep{r13} focuses on generating prediction intervals for $Y_{n+1}$, rather than parameter confidence sets.  Conformal prediction intervals assume data, $(Y_{i},X_{i}) \iid \mathbb{P},~i=1,\ldots,n$, are exchangeable and that the point estimate, $\hat{\mu}(X_{i})$, is symmetric.  Under these two assumptions, conformal inference outputs prediction intervals that achieve nominal coverage under any misspecification $\gamma$ for finite samples.  Remarkably, this result makes no assumptions about (the consistency of) $\hat{\mu}$ or distribution, $\mathbb{P}$ (though the resulting intervals will be relatively longer in proportion to the bias in $\hat{\mu}$).

\citet{r14} construct conformal prediction intervals under a Bayesian modeling setup.  The posterior predictive distribution, $g_{i} := g(Z_{1},\ldots,Z_{n+1};Z_{i}) = p(Y_{i} \mid X_{i},Z_{1},\ldots,Z_{n+1}) = \int f_{\theta}(Y_{i}\mid X_{i}) \times \pi(\theta \mid Z_{1},\ldots,Z_{n+1})d\theta$, where $Z_{i} = (X_{i},Y_{i})$.  In their set-up the posterior predictive distribution for each $i$ is used as the ``conforming score".  Score $g_{i}$ is exchangeable because the $\pi(\theta \mid Z_{1},\ldots,Z_{n+1})  ~\propto~  \pi(\theta) \times \mathop{\prod}_{i=1}^{n+1} f_{\theta}(Y_{i}\mid X_{i})$.

The resulting rank of $g_{n+1}$, $\pi(y) = \frac{1}{n+1}\mathop{\sum}_{i=1}^{n+1}\mathbf{1}(g_{i} \leq g_{n+1})$ is uniformly distributed such that one may form the conformal predictive set, $C_{\gamma}(X_{n+1}) = \left\{y \in \mathbb{R}: \pi(y) > \gamma\right\}$, where $y$ is sampled on a grid and the resulting intervals are guaranteed to achieve nominal coverage.  In the typical case the model must be re-fit for each value of $y$, though \citet{r14} devised an importance re-sampling scheme to re-weight the posterior draws for parameters, $\theta$, estimated from the first $n$ data points.

A key assumption to produce conformal prediction intervals under a Bayesian hierarchical model implementation (besides exchangeability) is that valid parameter samples from the model posterior distribution, $\pi(\theta \mid Z_{1},\ldots,Z_{n})$, are available.  In our set-up, however, we don't have a samples from a valid posterior distribution due to our use the computationally scalable ADVI estimation algorithm.  Even more, our focus is not on prediction of the next datum, but on estimation of the marginal parameter posterior distributions.

\citet{rChen} develop a simulation-based calibration procedure that is similar to ours, though their focus is on frequentist estimation of global parameters under variational inference.   They utilize a non-parametric bootstrap procedure since their inferential interest is on model \emph{global} parameters.  By contrast, our procedure uses a parametric-based re-sampling procedure because our inferential interest is in \emph{local} parameters, $\theta_{i}$, indexed by observation / domain (such that the re-sampling with replacement procedure for the non-parametric bootstrap would disallow inference on for local parameters indexed by domain $i$).   \citet{rChen} form a pivot statistic, but construct their variance estimate for each re-sampled dataset by using a sandwich estimator that requires analytical computation of the gradient and Hessian.   By contrast, we develop a numerical variance adjustment factor that is applied to the estimated variance, $v(\theta_{i})$, from the original run under the approximate estimation algorithm that does not require access to analytical gradients and Hessians.

The remainder of this paper presents our target model formulation in Section~\ref{sec:allmodels} that motivates our calibration adjustments of marginal posterior distributions.   We introduce our resampling-based calibration adjustment method in Section~\ref{sec:calibration} followed by simulation studies in Section~\ref{sec:simulations} that apply our calibration method to our models of 
Section~\ref{sec:allmodels}.  We subsequently apply our simulation-based calibration procedure to correct uncertainty quantification on a real data application from the Current Employment Statistics survey in Section~\ref{sec:application} followed by a concluding discussion in Section~\ref{sec:discussion}.

\section{Model Formulations to Apply Simulation-based Calibration}\label{sec:allmodels}

We introduce a domain-level model used in the Current Employment Statistics (CES) survey that estimates total employment in small domains defined by intersections of industry and geography.  This application motivates our development of the resampling-based calibration algorithm introduced in the next section that we subsequently use to correct bias in the mean and variance estimates from approximate marginal posterior distributions that we obtain from a computationally scalable optimization algorithm estimated on our domain-level model. 

The domain-indexed CES total employment point estimate, $y_{i}$, for domain $i \in 1,\ldots,N$, is constructed as a survey weighted statistic of underlying responses for sampled business establishments in that domain.  These survey-weighted domain statistics are referred to as direct estimates and are not model-based.  An accompanying statistic, $v_{i}$, representing the variance of $y_{i}$ (with respect to the distribution governing the taking of samples from a fixed population) is also constructed.  

Our inferential interest is extracting local parameters, $\theta_{i}$, that represent latent ``true'' domain point estimates.  Our model is a so-called small domain estimation formulation where correlations between domains with relatively large respondent sample sizes and those with relatively small respondent sample sizes are used to improve estimation quality in the latter (as compared to the direct point estimate, $y_{i}$) (See \citet{10.1214/22-AOAS1704} for a background on small domain estimation).

We developed and refined our total employment model for CES domains by estimating the model on CES employment historical series and subsequently testing its performance (e.g., mean-squared prediction error) against a ``gold standard'' of census/population-based employment levels that become available to researchers several months \emph{after} the actual publication of the CES estimates.  The results showed satisfactory performance of the model fitted point estimates, as compared to direct sample based estimates or estimates based on legacy CES models. 

We estimate our CES model using the Variational Bayes (VB) ADVI algorithm implemented in Stan. The VB algorithm is relatively fast, making it feasible for use under a tight production timeline to produce monthly survey-based domain point estimates and variances. 

Based on this research, we are satisfied with the model fitted point estimates of employment.   Our model framework specifies a joint distribution for domain-level point estimate and variance statistics.   The variance statistic measures the variation of the point estimate with respect to the distribution governing the taking of samples from a fixed population.   The variance statistic is noisy and our model produces a ``true" variance of the point estimate.   We co-model the point estimate and variance because they are correlated and because the variance provides information about the quality of the point estimate for regulating model-induced shrinkage.  That said, we are interested in the variance of the ``true" point estimate for each domain, $\theta_{i}$ (for domain $i \in 1,\ldots,N$), from the model, which is based on the model credibility intervals for $\theta_{i}$. 

Yet, the credibility intervals for the estimated true point estimate are based on draws from mean field variational algorithm that assumes the joint posterior distribution is fully factored.  So, we would \emph{a priori} expect the credibility intervals to be too optimistic and induce undercoverage. 

Our chosen model may be viewed as a generalization of the classical area-level Fay-Herriot (FH) model. Thus, we start by formulating the FH model (Section~\ref{sec:FH}) and discussing its shortcomings to motivate the extension to our model formulation. A modification of the FH model that allows co-modeling of direct point estimates and their variances is given in Section~\ref{sec:FHS}. Finally, we formulate the ``production version" of the model in Section~\ref{sec:models}.

\subsection{The Fay-Herriot model}~\label{sec:FH}
The Fay-Herriot (FH) model's name comes from \cite{r1} paper, where the authors adapted the James-Stein estimator for making estimates of income for small places. The hierarchical formulation of the model is also familiar to many from the classical eight schools example of \cite{r5} and it is used as an introductory example to hierarchical modeling by \cite{r2}.

Let ${y_i}$ be a survey estimate of target ``true" point estimate ${\theta }_{i}$ for domain $i$ (in the case of CES, it represents a relative change in employment from month $t-1$ to month $t$ in domain $i$); for each domain, $i=1,...,N$, assume
\begin{align}
{y_i} \color{black}{\vert\theta_i} & \ind {{N}\left( {\theta }_{i},{v_{i}} \right) \label{eq:FHlevel1}},\\	
{\theta }_{i}|\mathbf{\beta },\tau _{u}^{2} &\ind{N}\left(\mathbf{x}_{i}^{T}\mathbf{\beta },\tau _{u}^{2} \right). \label{eq:FHlevel2}
\end{align}	

Note that variances  ${v_{i}}, i=1,...,N$, in Equation~\ref{eq:FHlevel1} are assumed to be \emph{fixed and known}. 

In practice, true variances of survey estimates are not known, and $v_{i}$ represents some \emph{estimate} of the variance. Since direct survey estimates of variances contain noise, the traditional approach is to use some sort of \emph{preprocessing} to smooth variances (for example, based on a generalized variance function, or GVF) before they are used in the modeling.

\subsection{Co-modeling point estimates and their variances}~\label{sec:FHS}
A more efficient approach seeks to remove noise from the direct variance estimates by fitting them, simultaneously, with the point estimates in a single model (\cite{r4}, \cite{r6}, \cite{r3}). For each domain $i$ we consider data vector ${\left( {{y}_{i}},{{v}_{i}} \right)}$ of direct survey estimates. Assume the following model for $i=1,...,N$:
\begin{align} 				
{y_i}  \color{black}{\vert\theta_i},{\sigma _{i}^{2}} &\ind N\left( {{\theta }_{i}}, {\sigma _{i}^{2}} \right)\label{eq:FHSlevel1}, \\				
{{\theta }_{i}}|\mathbf{\beta },\tau _{u}^{2} &\ind N\left( \mathbf{x}_{i}^{T}\mathbf{\beta },\tau _{u}^{2} \right)\label{eq:FHSlevel2}.	\\			
{v_{i}}\color{black}{\vert a},{\sigma _{i}^{2}} &\ind G\left( \frac{an_{i}^{*}}{2},\frac{an_{i}^{*}}{2{\sigma _{i}^{2}}} \right),\label{eq:FHSlevel3} \\							
{\sigma _{i}^{2}}\vert\mathbf{\gamma } &\ind IG\left( 2,\exp \left( z_{i}^{T}\mathbf{\gamma } \right) \right),	\label{eq:FHSlevel4}	
\end{align}	

where $G(\alpha,\beta)$ and $IG(\alpha,\beta)$ denote the gamma and inverse gamma distributions with shape parameter, $\alpha$ and scale parameter $\beta$, respectively. 

The shape and scale parameters of the gamma distribution depend on an \emph{unknown} parameter $a$ and domain sample size $n_{i}$ (we use standardized values for sample sizes, \newline $n_{i}^{*}={\left( n_{i}-\left\{ \underset{i}{\mathop{\min }}\,n_{i}-1 \right\} \right)}/{\left( \underset{i}{\mathop{\max }}\,n_{i}-\underset{i}{\mathop{\min }}\,n_{i} \right)}\;\in \left[ 0,1 \right]$). \\

Note that $a n_{i}^{*}$ controls the variation in the likelihood around latent true variance ${\sigma _{i}^{2}}$, where domains with relatively larger values for $n_{i}^{*}$ will have smaller variation, which accords with the notion that the observed variances ${v_{i}}$ are of relatively higher quality for domains with larger sample. Parameter $a$ allows the data to estimate a uniform scale of this sample size effect.  Equation~\ref{eq:FHSlevel4} denotes the prior distribution for the true variance ${\sigma _{i}^{2}}$; the true variance has an inverse gamma prior distribution where the mean depends on vector of covariates $z_{i}$ through $\exp \left( z_{i}^{T}\mathbf{\gamma } \right)$, where $\mathbf{\gamma }$ are unknown parameters.  We label this model as ``FHV''.

\subsection{Relaxing linearity assumptions using finite mixtures}~\label{sec:models}
We now introduce our main model from \cite{rS1}. Here, we assume that Equation~\ref{eq:FHSlevel1} holds and relax the parametric assumptions of Equations~\ref{eq:FHSlevel2},~\ref{eq:FHSlevel3}, and~\ref{eq:FHSlevel4} by specifying finite mixtures in place of each of their respective distributions,
\begin{align} 				
{{\theta }_{i}}|\mathbf{\pi },\mathbf{\mu },\mathbf{\beta },\tau _{u}^{2} &\iid \sum\nolimits_{k=1}^{K}{{{\pi }_{k}}N\left( {{\mu }_{k}}+\mathbf{x}_{i}^{T}\mathbf{\beta },\tau _{u}^{2} \right)} \label{eq:CFHGlevel2}\\				
v_{i}^{{}}|a,\sigma _{i}^{2},\mathbf{b},\mathbf{\pi } &\ind\sum\nolimits_{k=1}^{K}{{{\pi }_{k}}}G\left( \frac{an_{i}^{*}}{2},\frac{an_{i}^{*}}{2{{b}_{k}}\sigma _{i}^{2}} \right) \label{eq:CFHGlevel3} \\			 			
\sigma _{i}^{2}|\mathbf{\gamma },\mathbf{\pi } &\ind\sum\nolimits_{k=1}^{K}{{{\pi }_{k}}IG\left( 2,\exp \left( z_{i}^{T}{{\mathbf{\gamma }}_{k}} \right) \right)} \label{eq:CFHGlevel4}.
\end{align}		

The maximum number of clusters, $K$, is over-specified and denotes a truncated model.  The actual number of discovered clusters, $K^{\ast} \leq K$, is governed by the following prior formulations,
\begin{align}
\pi_{k}\vert \alpha &\sim D\left(\frac{\alpha}{K},\ldots,\frac{\alpha}{K}\right) \label{eq:Dir}\\
\alpha &\sim G(a_{\alpha}=1,b_{\alpha} = 1)\\
\gamma_{k} &\ind N_{P}(\mu_{\gamma} = 0,\Sigma_{\gamma} = \mathbb{I}_{P})\\
b_{k} &\ind G(\frac{\beta_{b}}{2}-1,\frac{\beta_b}{2}),
\end{align}
where $D(\cdot)$ denotes the Dirichlet distribution, and $\mathbb{I}_{P}$ denotes the $P\times P$ identity matrix. Equation \ref{eq:Dir} encourages sparsity in the number of clusters/components discovered and makes the setup equivalent to a truncated Dirichlet process.  We set hyperparameter, $\beta_b = 4$ in the prior for $b_{k}$ and $P$ denotes the number of predictors in $z_{i}$.  \\

The formulation is completed by utilizing the specification of a matrix of continuous predictors that allows the prior distributions for any two domains to cluster together to be influenced by predictors, $\mathbf{g}_{i}^{{}}$, so that two domains, ${{i}_{1}}$ and ${{i}_{2}}$, that have similar predictor values, $\mathbf{g}_{{{i}_{1}}}^{{}}$ and $\mathbf{g}_{{{i}_{2}}}^{{}}$ would be assigned a higher probability to cluster together, \emph{a priori}.  We accomplish this setup by placing a further prior distribution on predictors, $\mathbf{g}_{i}$ with,
\begin{equation} \label{eq:CFHGlevel5}
\mathbf{g}_{i}^{{}}|\mathbf{\mu }_{g}^{{}},\mathbf{\Sigma }_{g}^{{}},\mathbf{\pi }\ind\sum\nolimits_{k=1}^{K}{{{\pi }_{k}}{{\mathbf{N}}_{l}}\left( \mathbf{\mu }_{gk}^{{}},\mathbf{\Sigma }_{gk}^{{}} \right)},
\end{equation}

where we influence the co-clustering of domains by incorporating mixture parameters, $\left( \mathbf{\mu }_{gk}^{{}}, \mathbf{\Sigma }_{gk}^{{}} \right)$, into the set of clustered parameters. Using predictors to influence the prior probability of co-clustering parameterizes a predictor-dependent clustering formulation and is another way to use predictors beyond a regression relationship with the response variable.  We do \emph{not} believe the predictors, $\mathbf{g}_{i}$, are random and are not interested in the fitted values of parameters $\mathbf{\mu }_{gk},\mathbf{\Sigma }_{gk}^{{}}$, but fitting $\mathbf{g}_{i}^{{}}$ and including the $\mathbf{\mu }_{gk}^{{}}$ and $\mathbf{\Sigma }_{gk}$ in the clusters along with ${{\mu }_{k}}$, $b_{k},{{\mathbf{\gamma }}_{k}}$ helps better identify the clustering structure to give sharper estimates.  Placing a probability model on $\mathbf{g}_{i}$ is a device that allows the use of the simple truncated mixture to parameterize a predictor-dependent clustering model \citep{muller2011}.  

Finally, parameters $b_{k}$ allow the data to estimate a bias in the observed variances $v_{i}$ (for those domains $i$ linked to cluster $k$).  The bias $b_{k}$ multiplicatively offsets the estimated true variance, $\sigma_{i}^{2}$.  (See \citet{rS1} for more details.)  We label our resulting model with ``CFHV" to denote a clustering version of the Fay-Herriot that also co-models the variance.

\section{Resampling-based Calibration}\label{sec:calibration}

We use the ADVI estimation algorithm (\citet{r11}) implemented in Stan to estimate the mixture model described in Section \ref{sec:models}. This algorithm implements a mean field variational Bayes approximation for the posterior distribution characterized by a fully factorized normal distribution (on a transformed parameter space such that each transformed parameter lies on the real line).  It has been reported in the literature (\citet{r12}) that the mean field approximation may produce inadequate measures of uncertainty for model parameters, even if the resulting posterior means are unbiased estimates of parameters.

In this section we consider a re-sampling, bootstrap-like procedure (that we will refer to as the ``bootstrap" procedure in the sequel) to evaluate and re-calibrate potential
bias in the second moment of the estimation algorithm that improves the estimation accuracy of parameter variances.  BLS publish both point and variance estimates for CES domains, where each variance connotes a measure of estimation quality and its publication is facilitated by our proposed re-calibration procedure.   We illustrate how to use our re-calibration procedure to also correct bias in the first moment, though use of approximate estimation algorithms like ADVI are typically predicated on producing high quality point estimates (from first moments).  We further demonstrate how to use re-calibrated first and second moments to adjust confidence intervals that ensure the nominal coverage at any level of misspecification $\gamma$. 

The bootstrap procedure is now described in the following sequence of steps:

\begin{enumerate}
    \item Estimate a hierarchical model (e.g., of Section~\ref{sec:models}) using the ADVI algorithm on the given data.  Let $m(\theta_{i})$ and $v(\theta_{i})$ denote the approximate posterior mean and variance, respectively, of parameter $\theta_{i}$ for domain $i$ obtained from the initial model estimation using ADVI.  Extract random \emph{draws} of model parameters ${\theta}_i^{(\alpha)}$ indexed by $\alpha = (1,\ldots,A)$ from the approximate posterior distribution of model parameters ${\theta}_i$ from the initial model run with ADVI.  Use these draws, ${\theta}_i^{(\alpha)}$ to, in turn, generate draws of replicate data ${y}_i^{(\alpha)}$ from the approximate posterior predictive distribution of ${y}_i$ conditioned on the parameter draws. This procedure constructs $A$ re-sampled datasets for each domain, $i \in 1,\ldots,n$. 
    
    In each dataset, obtained from draw $\alpha$, we treat the associated  ${y}_i^{(\alpha)}$ as ``observed" data, and ${\theta}_i^{(\alpha)}$ as ``bootstrap true'' parameters for $i = (1,\ldots,n)$.
    
    \item Estimate the model \emph{again} under the same ADVI algorithm for \emph{each} re-sampled dataset $(\mathop{{y}_i^{(\alpha)}})_{i=1}^{n}$, indexed by $\alpha=1,\ldots,A$. We obtain approximate posterior means, $m({\theta}_i^{(\alpha)})$ and variances, $v({\theta}_i^{(\alpha)})$ for respective parameters from each model estimation on dataset $\alpha$.

    \item Use model estimations on the $A$ re-sampled datasets to adjust the first moment, as follows: 
        
        The bias of the model fitted approximate posterior mean $m({\theta}_i)$ of parameter $\theta_i$ is expressed with,
        \begin{equation*}
            {a_i}={m({\theta}_i)}-{\bar{m}(\theta_i^{(\alpha)})},
        \end{equation*}
        where $m({\theta}_i)$ is the approximate posterior mean from the initial run and $\bar{m}({\theta}_i^{(\alpha)})={A}^{-1}\sum\nolimits_{\alpha=1}^{{A}}m({\theta}_i^{(\alpha)})$ is the bootstrap mean, obtained as the average of approximate posterior means, $m({\theta}_i^{(\alpha)})$, over the $A$ replicate runs. The first order bias adjusted posterior mean for ${\theta}_i$ is
        \begin{equation}
            \Tilde{m}(\theta_i)=m({\theta}_i)+{a_i}.
        \end{equation}
       
    \item Use the $A$ model estimations to formulate a second moment adjustment that we next derive from properties of the pivot statistic,
    \begin{equation}\label{eq:pivot}
                T_i=\frac{{m}({\theta}_i)-{\theta}_i}{\sqrt{{v}({\theta}_i)}}. 
    \end{equation}
    If ${m}({\theta}_i)$ and ${v}({\theta}_i)$ are consistent estimates of the domain point estimate and variance, respectively, then the pivot is distributed as a t-statistic with $(0,1)$ mean and variance, respectively, that contracts on a standard normal distribution.  We write that $T_{i} \sim (0,1)$ as a shorthand notation that denotes this statistic is distributed under a symmetric distribution with mean $0$ and variance $1$. We next use this standardization of the pivot statistic to derive our adjustment to the estimated variance.

    We know that ${v}({\theta}_i)$ is \emph{not} a consistent estimator of the true variance (of the numerator of the pivot) when estimated from the ADVI algorithm.   We define variance scale adjustment, $c_{i}$, under an assumption that the variance correction factor is of a multiplicative scaling formulation,
    \begin{align*}
    \Var(\theta_i)=v(\theta_i)c_i,
    \end{align*}
    where $\Var(\theta_i)$ denote the true (posterior) variance of the numerator in Equation~\ref{eq:pivot}.

    We estimate the pivot statistic of Equation~\ref{eq:pivot} from its distribution over the $A$ bootstrap samples (each indexed by $\alpha \in 1,\ldots,A$) using, 
    \begin{equation}\label{eq:bootpivot}
                T_i^{(\alpha)}=\frac{{m}({\theta}_i^{(\alpha)})-{\theta}_i^{(\alpha)}}{\sqrt{{v}({\theta}_i^{(\alpha)})}}. 
    \end{equation}
    \begin{enumerate}
    \itemsep 3mm
     \item We first focus on correcting or adjusting variance of the bootstrap runs, $v(\theta_i^{\alpha})$, where the variance of the original model run, $v(\theta_i)$, is treated as the truth.  The bootstrap variances may be corrected back to the original model run using, $v(\theta_i)=v(\theta_i^{\alpha})c_i$.  Thus, we divide pivot $T_i^{\alpha}$ by $\sqrt{c_i}$ to correct the bootstrap variances in the pivot statistic back to the bootstrap truth.
    \item However, since we draw our bootstrap samples from an approximate posterior distribution with \emph{biased} variance $v(\theta_i)$, we must again divide pivot $T_i^{\alpha}$ by $\sqrt{c_i}$, to correct back to the \emph{true} posterior variance, $\Var(\theta_i)$.
    \end{enumerate}
    We use the standardization property of the pivot under use of the correct variance to derive our adjustment from,
    \begin{equation*}
        \Var(c_i^{-1}T_i^{(\alpha)})=1. 
    \end{equation*}
    which gives us $c_i^2=\Var(T_i^{(\alpha)})$ for the form of $c_{i}$.  This form for $c_{i}$, in turn, produces, ${c}_{i} =\sqrt{A^{-1}\sum\nolimits_{\alpha=1}^{{A}}({T}_i^{(\alpha)}-\bar{T_i})^2}$, where $\bar{T_i}=A^{-1}\sum\nolimits_{\alpha=1}^{{A}}{T}_i^{(\alpha)}$.
    
     \item \emph{Adjusted confidence intervals (II): the pivotal approach} 

     The adjusted pivot after performing the variance correction using the estimated $c_{i}$ in the previous step is,
          \begin{equation*}
           \Tilde{T}_i^{(\alpha)}=\frac{{T}_i^{(\alpha)}-\bar{T_i}}{c_i} \sim (0,1),
           \end{equation*}

        which we invert to produce the calibrated level-$\gamma$ confidence intervals.
        \begin{align}\label{eq:pivotinterval}
                C_i(\gamma)=\left[\Tilde{m}({\theta}_i)  + \sqrt{{v}({\theta}_i){{c}_{i}}}{\Tilde{t}_{i,\gamma}},
                \Tilde{m}({\theta}_i)  + \sqrt{{v}({\theta}_i){{c}_{i}}}{\Tilde{t}_{i,1-\gamma}}\right],
            \end{align}
        
        \item \emph{Adjusted confidence intervals (II): direct re-scaling approach.} We can obtain adjusted confidence intervals for fitted ${\theta}_i$ by rescaling each draw ${\theta}_i^*$ from the approximate posterior distribution of ${\theta}_i$ produced by the ADVI algorithm to match the first two moments:
        \begin{equation}
            \Tilde{\theta}_i^{*}=\frac{\theta_i^{*}-m({\theta}_i)}{\sqrt{v(\theta_i)}}\sqrt{{v}({\theta}_i){c_i}}+\Tilde{m}({\theta}_i),
        \end{equation}  
        where we use the same $c_{i}$ derived from standardizing the bootstrap indexed pivot to scale adjust the approximate posterior variance, ${v}({\theta}_i)$. The adjusted confidence intervals are obtained by computing percentiles over the adjusted draws $\Tilde{\theta}_i^*$.
\end{enumerate}

\section{Simulation study}\label{sec:simulations}

\subsection{Calibration under the Fay Herriott (FH) Model}

In the first part of our simulation study, we generate data from the relatively simple Fay-Herriot model and then fit the data using ADVI, evaluate and adjust the variance to produce revised confidence intervals using the two methods to construct intervals proposed in Section \ref{sec:calibration}. We consider $N=150$ domains and let true model parameters be $\beta=1$, $\tau_u^2=1$, $\sigma_i^2=1$; we generate covariates as $x_i \ind Unif(0,2)$, random effects as $u_i \ind N\left( 0, \tau_u^2 \right)$, and random errors as $\epsilon_i \ind N\left( 0, \sigma_i^2 \right).$ True domain values are  $\theta_i=\mathbf{x}_{i}^{T}\mathbf{\beta }+u_i$ and ``direct'' domain estimates are $y_i=\theta_i+\epsilon_i.$ For this setup, variances of $y_i$ are assumed to be observed and measured exactly ($v_i=\sigma_i=1.)$ 

We run the model and extract $S=200$ \emph{simulation} datasets from the posterior distribution of $\theta_i$ and posterior predictive distribution of $y_i$, where $S$ denotes the number of Monte Carlo simulation iterations (that allows us to assess coverage of our rescaled estimator). Next, we perform the re-sampling algorithm $A=500$ times within \emph{each} of the $S=200$ Monte Carlo simulation datasets.

\begin{table}[h]
\centering
\caption{Coverage properties for model fitted $m({\theta}_i)$,  50\% nominal, over 150 domains and $S=200$ simulation runs, using $A=500$ re-samples
}
\label{tab:simFH_covtable}
\begin{tabular}{c|ccc}
   & Orig Fitted & Rescaled & Pivot  \\
\hline
 Coverage  & 0.531 & 0.493& 0.492\\
 Length  & 1.019 & 0.935& 0.933\end{tabular}
\end{table}

To assess the coverage properties of the confidence intervals (CIs), we compute the number of times, over $S=200$, that the CI - before and after adjustment - covers the true parameter value. We implement the pivotal and rescaling based adjustments solely to adjust the second moment and not the first moment of the model estimation algorithm.  So, we use the posterior means, $m(\theta^{(\alpha)}_{i})$ (in place of $\Tilde m(\theta^{(\alpha)}_{i})$) to form revised confidence intervals.  Our experience reveals that the ADVI produces accurate point estimates for our CES models of Section~\ref{sec:models}.  Including the first order bias correction in this case where ADVI produces relatively accurate point estimates induces added noise in the calibrated confidence intervals.

Results for 50\% nominal coverage, averaged over all 150 domains for $S=200$ simulation runs, are presented in Table \ref{tab:simFH_covtable}. The original fit produces slight overcoverage; both rescaled and pivotal based adjustment methods give nearly nominal coverages. The lengths of the adjusted intervals are shorter than the original lengths. 

\begin{figure}[ht!]
  \begin{center}
  \includegraphics[width=0.7\linewidth]{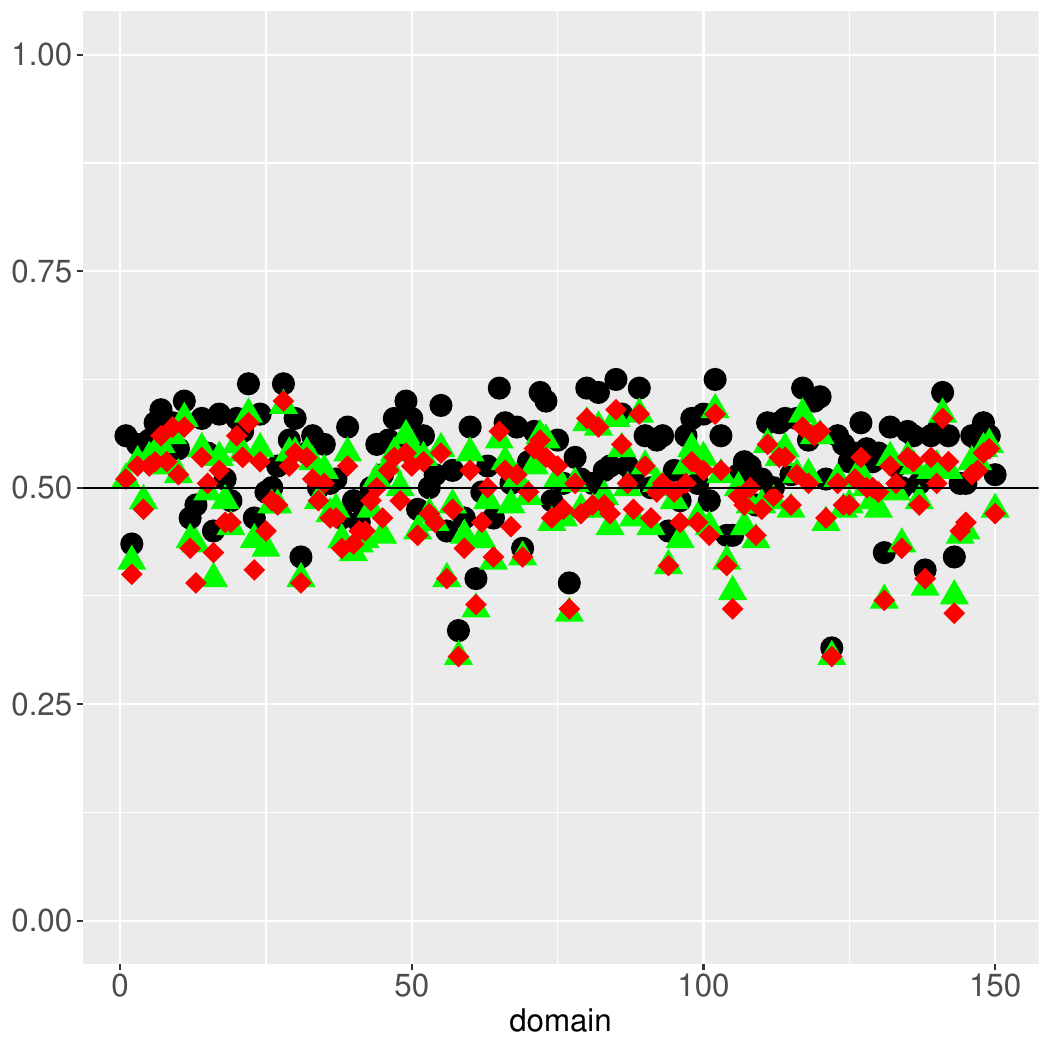}
  \caption{Domain coverages (50\% nominal) of fitted point estimates vs number of respondents. Black dots represent originally fitted model; green triangles represent adjusted coverages using rescaling method; red diamonds represent adjusted coverages based on the pivotal method.}
  \label{fig:simFH_cov}
  \end{center}
\end{figure}

In Figure \ref{fig:simFH_cov}, we plot coverages for \emph{each} domain. The black dots represent coverages from the original fit, the green triangles and red diamonds correspond to the rescaled and pivotal based adjustment methods, respectively. The adjustments bring coverages closer to the nominal 50\% line.    

\begin{figure}[ht!]
\begin{center}
  \includegraphics[width=0.65\linewidth]{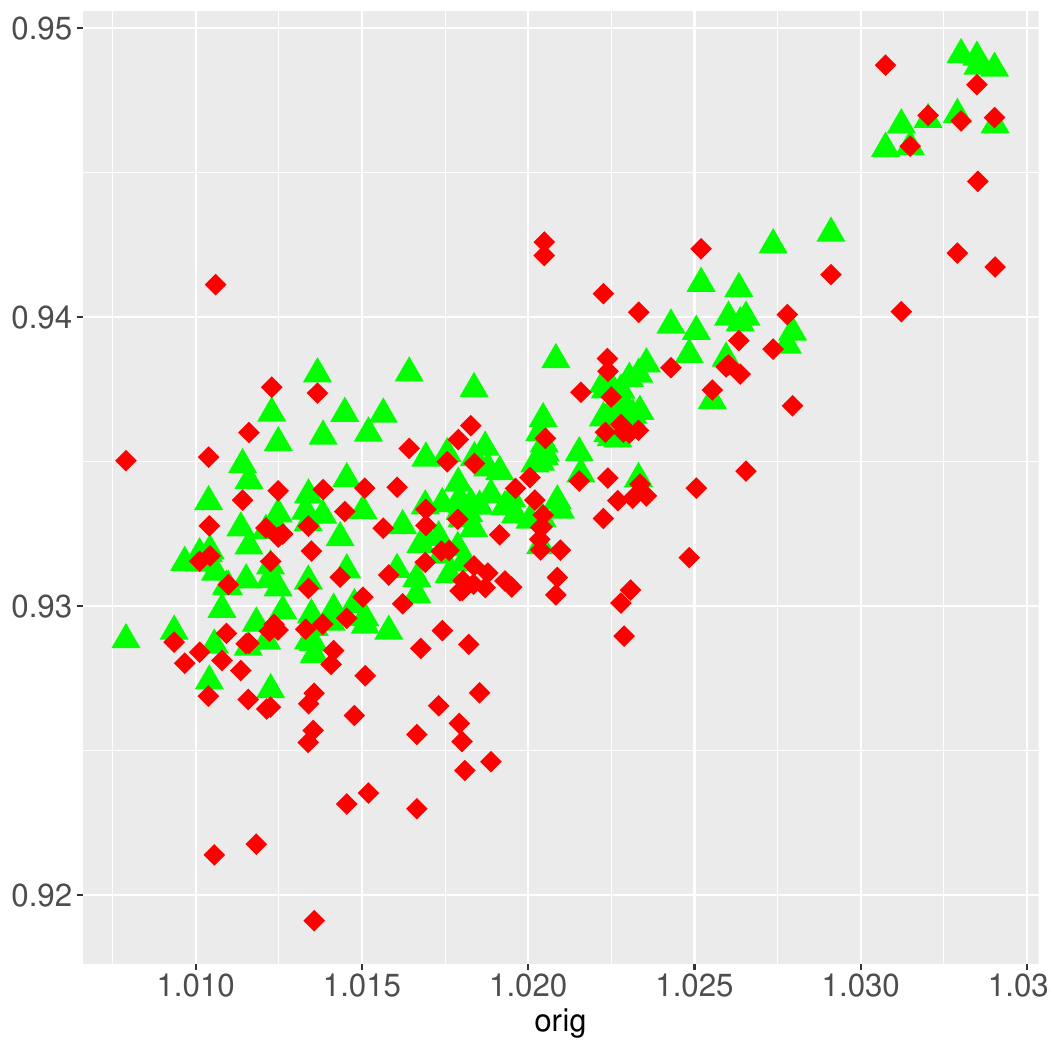}
  \caption{Lengths of the adjusted vs the original (50\% nominal) confidence intervals. Black dots represent originally fitted model; green triangles represent adjusted coverages using rescaling method; red diamonds represent adjusted coverages based on the pivotal method.}
  \label{fig:simFH_length_both}
\end{center}
\end{figure}

In Figure \ref{fig:simFH_length_both}, we plot the length of the adjusted CIs against the lengths of the original CIs. The lengths of CIs from the rescaled and pivotal based adjutment methods (green triangles and red diamonds, respectively) are shorter than the original CIs. The pivotal based CIs are somewhat noisier but a little shorter than the rescaled-based CIs.

\subsection{Calibration under the clustering CFHV model}
In the second part of our simulation study, we evaluate the proposed calibration procedure for a more complicated CFHV model, given in Section \ref{sec:models}.  For this part of the study, simulation datasets are generated as follows: We first estimate the mixture model of Section~\ref{sec:models} on an available \emph{real} data set and extract $S=200$ random draws from the posterior distribution of model parameters. These can be viewed as ``true" values of the model parameters for dataset $s$, $s=1,\ldots S$. We use the same initial run to draw ``observed" data $({y}_i^{(s)},{v}_i^{(s)})$ from the posterior predictive distribution of $({y}_i,{v}_i).$ These data, along with covariates, domain labels, number of respondents from the original dataset form our $S$ simulated datasets. As noted, the dataset is based on real domains of various sizes with the number of respondents ranging from 1 to 247 over the total of 166 domains.

For each simulation dataset $s$, we generate $A=500$ re-sampled datasets obtained from a model estimation run and perform the calibration procedure as described in Section \ref{sec:calibration}.  

We use posterior draws from respective $A = 500$ model estimation runs (for each dataset, $s \in 1,\ldots, S$) to construct CIs for each model parameter and compute the number of times, over $S=200$, that the CI - before and after adjustment - covers the true parameter value.   

We present coverage properties (for 50\% nominal coverage) of the confidence intervals for model fitted $m({\theta}_i)$ in Table \ref{tab:sim_covtable}. The results in the table are averaged over all domains.  One notices the overcoverage of the original (``Orig Fitted") intervals obrained from the model posterior draws. The adjustment using the rescaling and pivotalbased approaches brings the coverage closer to the nominal.

\begin{table}[h]
\centering
\caption{Coverage properties for model fitted $m({\theta}_i)$,  50\% nominal, over 166 domains and $S=200$ simulation runs, using $A=500$ re-samples
}
\label{tab:sim_covtable}
\begin{tabular}{c|ccc}
   & Orig Fitted & Rescaled & Pivot   \\
\hline
 Coverage  & 0.625 & 0.559& 0.549\\
 Length  & 0.618 & 0.537& 0.528\end{tabular}
\end{table}

\begin{figure}[h!]
\centering
  \includegraphics[width=0.65\linewidth]{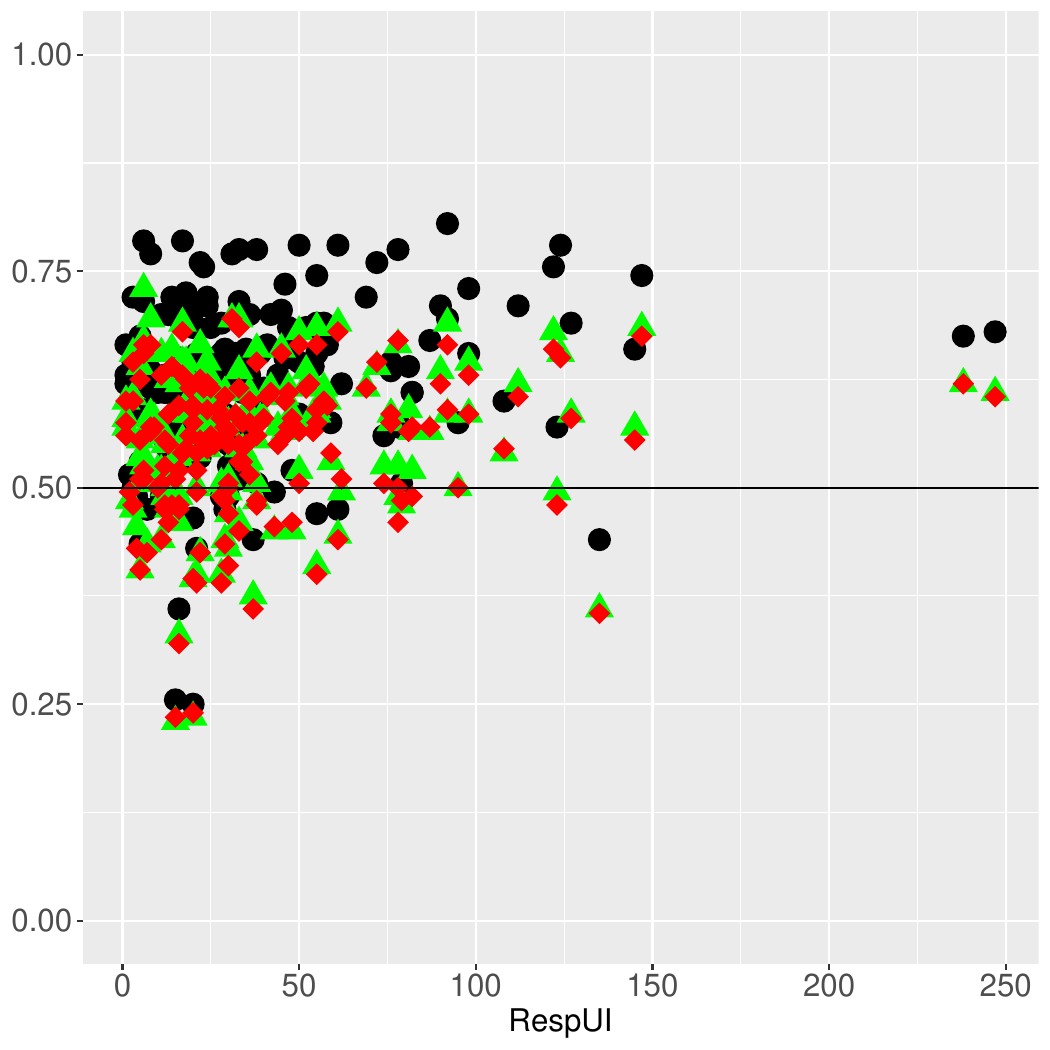}
  \caption{Domain coverages (50\% nominal) of fitted point estimates vs number of respondents. Black dots represent originally fitted model; green triangles represent adjusted coverages using rescaling method; red diamonds represent adjusted coverages based on the pivotal method. Domains are ordered along the horizontal axis by increasing number of sampled establishments from left-to-right.}
  \label{fig:sim_coverage_both}
\end{figure}

To complement the overall results presented in Table \ref{tab:sim_covtable}, we present plots showing the distribution of the coverages over domains. We show coverage properties for \emph{each} domain of the $50\%$ confidence intervals in Figure \ref{fig:sim_coverage_both} computed over the $S$ simulation runs - both before adjustment and after adjustment under the rescaling and pivotal based approaches. Coverage for point estimates is plotted on the vertical axis vs the number of respondents in a given domain, on the horizontal axis where the domains are ordered according to an increasing number of sampled establishments (sampled in a domain) from left-to-right. Black dots represent the original model results, green triangles show results from the rescaling method, and red diamonds represent the pivotal based adjusted coverages. We observe overcoverage of the original model fit for most of domains. The rescaling and pivotal based approaches produce similar results for dispersion of domain-indexed coverages around the truth $50\%$ nominal, though the pivotal based coverage results are slightly more compactly distributed along the horizontal line representing the nominal $50\%$ coverage. 

\FloatBarrier

\begin{figure}[h!]
\centering
  \includegraphics[width=0.7\linewidth]{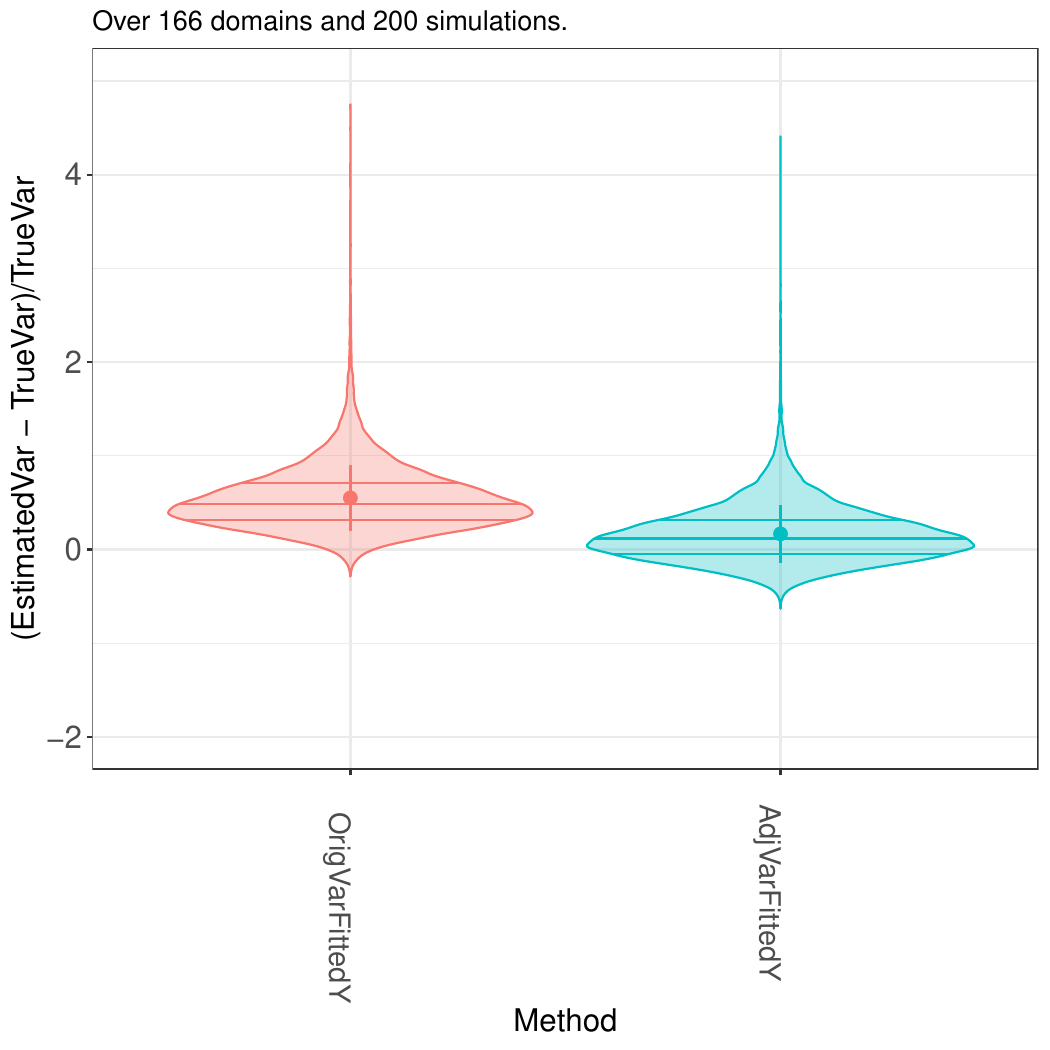}
  \caption{Distributions, over 166 domains and 200 simulation runs, of residuals of model posterior variances $v(\theta_i)$ and adjusted variances $\Tilde{v}(\theta_i)$, relative to respective true variances of $\theta_i$. 
  ("Original" - model approximate posterior variances; "Adjusted" - adjusted variances).}
  \label{fig:sim_relvfitted_adj}
\end{figure}
\FloatBarrier

The violin plots in Figure \ref{fig:sim_relvfitted_adj} show distributions over domains and simulation runs of relative residuals of $v(\theta_i)$, the model approximate posterior variances of $\theta_i$, and also the adjusted variances $\Tilde{v}({\theta}_i)$. The original model posterior variances of $\theta_i$ are upwardly biased and the bias is corrected in the calibrated version of the variances.   

\begin{figure}
\centering
  \includegraphics[width=0.7\linewidth]{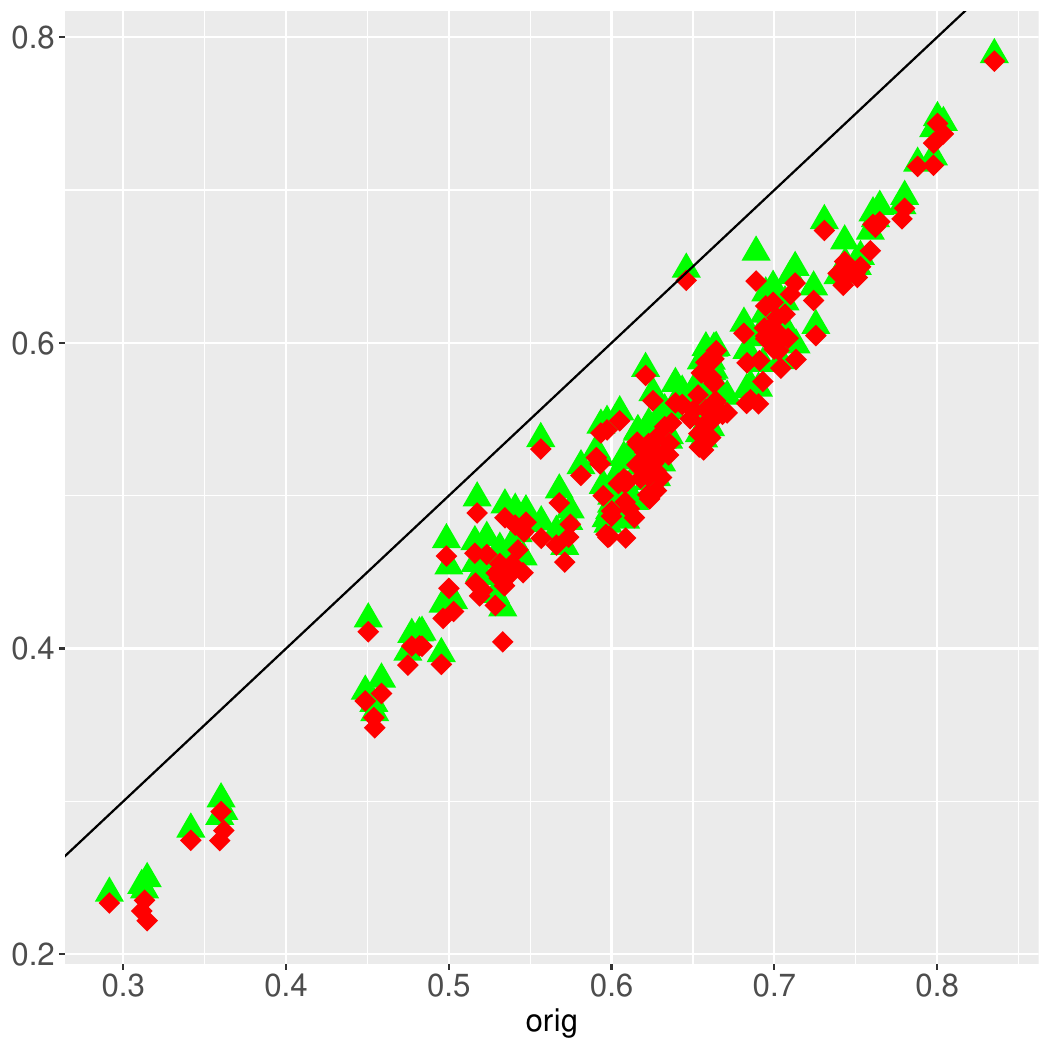}
  \caption{Lengths of the adjusted vs the original (50\% nominal) confidence intervals. Black dots represent originally fitted model; green triangles represent adjusted coverages using rescaling method; red diamonds represent adjusted coverages based on the pivotal method.}
  \label{fig:sim_length_both}
\end{figure}

In Figure \ref{fig:sim_length_both}, we plot lengths of the adjusted confidence intervals against the lengths of the original CIs. Each symbol corresponds a domain. The adjusted intervals are shorter than the original ones, with the pivotal based mathod being slightly shorter than the rescaling based.

\FloatBarrier

\section{Convergence of Adjusted Uncertainty Intervals}\label{sec:convergence}

We next discuss a convergence result of our corrected $\gamma-$level uncertainty intervals to nominal coverage at $\mathcal{O}(N^{-1})$ rate where we recall that $N$ denotes the number of domains.  We leverage existing theoretical results from the bootstrap literature to assess the convergence properties of uncertainty intervals produced from our simulation-based calibration procedure.   To leverage these results, we require two conditions:  Firstly, the specified \emph{model} must be asymptotically correct for data generation because our simulation-based calibration procedure uses a parametric bootstrap since our interest focuses on inference for model parameters.  (See \citet{ghosal2017} for assumptions that guarantee consistency of a Bayesian model.  Two important assumptions are: 1. Bounded entropy that regulates the complexity of the model; 2. The joint prior distribution is required to place positive mass on the truth). Secondly, we require first-order consistency of the \emph{estimation} algorithm, though the converged value may express a constant or fixed bias (e.g., If the (unknown) true value for parameter $\theta_{i}$ is denoted by $\theta_{0,i}$, the estimating algorithm may contract onto fixed point, $\theta_{a,i} \neq \theta_{0,i}$).

We compare convergence of the adjusted marginal confidence interval of our pivotal method that is based on the re-sampling distribution to the scale correction method that performs a second order correction to the approximate posterior distribution to construct the marginal credibility interval.


We formulate a pivot statistic for each re-sampled dataset, $\alpha \in (1,\ldots,A)$ with scale adjustment factor $c_{i}^{2}$ multiplying the model posterior variance, $v({\theta}^{(\alpha)}_{i})$ in the denominator of the pivot indexed by re-sample iteration $\alpha$.  The quantiles of the resulting distribution of the pivot statistic over the re-sampled iterations are used to form a confidence interval for $\theta_{i}$ under our pivotal method.  So, these intervals are based on the re-sampling/empirical distribution (of the pivot statistic), not the approximate model posterior distribution (of $\theta_{i}$) obtained from the scalable estimation algorithm, such as ADVI.  We replace the approximate posterior distribution with the empirical distribution to formulate a calibrated confidence interval for $\theta_{i}$.  

\citet{Efron1993} discuss that the convergence of the pivotal method intervals under a bootstrap re-sampling procedure are ``second order accurate"; that is $\mathbb{P}\left(\theta_{i} \in C_{i}(\gamma)\right) \rightarrow 1-\gamma$ at an $\mathcal{O}(1/N)$ rate.  Our re-sampling procedure that draws replicate data from the model posterior predictive distribution may be viewed as equivalent to the bootstrap procedure conditioned on the correctness of the underlying model used to estimate the posterior predictive distribution.

\citet{Efron1993} notes that the variance of the distribution of the pivot statistic (that we will refer to as ``pivot distribution") is independent of $\theta_{i}$ such one does not have to estimate the variance by plugging in $\hat{\theta}_{i}$.  Therefore, the variance of the pivot distribution converges to the truth at a relatively faster (higher order) $\mathcal{O}(1/N)$ rate (versus the slower $\mathcal{O}(1/\sqrt{N})$ rate for the percentiles of the underlying bootstrap distribution that require plugging in $\hat{\theta}_{i}$, which converges to $\theta_{i}$ at $\mathcal{O}(1/\sqrt{N})$).  The convergence of the higher distribution moments (e.g., skewness and kurtosis) are $O(1/N^{3/2})$ and $\mathcal{O}(1/N)$, respectively.  Therefore, the resulting pivotal based confidence intervals converge to nominal coverage at a second order rate without requiring convergence to a normal distribution.  So, this rate will be faster than the convergence of intervals that require convergence of the estimated distribution to a normal distribution for intervals to achieve nominal coverage.  \citet{rChen} demonstrates this same result for intervals based on the pivot statistic when using variational inference to obtain point estimates for $\theta_{i}$.

Our re-scaling option, by contrast with the pivotal method, uses the variance correction factor, $c_{i}$, to directly perform a re-scaling (variance correction) of the draws taken from the approximate posterior distribution for $\theta_{i}$.  The convergence of the adjusted posterior distribution to the truth (required for convergence of the adjusted credibility intervals to achieve nominal coverage) requires that the adjusted posterior distribution converges to a normal distribution.  Even where the adjusted posterior distribution converges to a normal distribution, we expect the convergence of the associated credibility intervals to be at a slower first order ($\mathcal{O}(1/\sqrt{N})$) rate.  

The faster convergence of the pivotal based confidence intervals as compared to the re-scaled credibility intervals explains why their domain-indexed ($i \in (1,\ldots,N)$) coverages are more tightly clustered around the $1-\gamma = 50\%$ target for the former as compared to the those of the latter in Figure~\ref{fig:sim_coverage_both}.

Our motivating application on estimation of CES point estimates, $\theta_{i}$, determines our choice of the ADVI (mean-field variational Bayes) algorithm estimation.  BLS use ADVI to computationally scale the model estimations to support the monthly production process for CES employment estimates.  Using the re-scaling option to calibrate the uncertainty intervals requires that the underlying ADVI produces approximate posterior distributions that converge to a normal distribution.   \citet{10.5555/1036843.1036913} show that the mean field variational Bayes approximate posterior distribution \emph{does} achieve asymptotic normality for parametric models of the exponential family.   Yet, \citet{10.1214/19-AOS1883} demonstrate that the distribution error has two parts:  1. Convergence of the true posterior distribution; 2. Variational Bayes error.  For nonparametric and high dimensional models, a specific prior formulation must be used such that the first error source dominates the second and asymptotic normality is achieved.  If a prior distribution of this specific form is not used, however, the approximate posterior distributions will not converge to a normal distribution.  In this case, the second order correction will not resolve errors in higher moments and the resulting adjusted credibility intervals will not achieve nominal coverage, no matter how large is the sample size.

Even though it may be the case that most model formulations will produce approximate posterior distributions that converge to normal distributions under use of mean-field variational Bayes, there is yet some possibility that such will not the case for a specific model formulation suggests that one may not universally rely on uncertainty intervals formed from the adjusted (calibrated) posterior distribution.  Replacing the approximate posterior distribution with use of the empirical distribution over the re-samples avoids this issue of convergence to normality that is tied to the approximation algorithm and may be universally relied on.   It bears mention that there is a computational cost to re-estimate the model for each re-sampled dataset and there is some minimum (e.g., $A = 100$) re-samples required to control Monte Carlo estimation error to estimate the empirical distribution over the re-sampled parameter estimates (though the replicate model re-estimations may be parallelized if computational resources permit).

Lastly, our simulation-based calibration procedure may be used with \emph{any} estimation algorithm that propagates its error over multiple data estimations to calibrate confidence intervals that achieve nominal coverage.  While we have focused on ADVI, there are other computationally scalable algorithms that may be selected to estimate an approximation to the model posterior distribution \citep{JMLR:v21:18-817}.  Convergence of the approximate posterior distributions to (true) normal distributions would have to be verified for each estimation algorithm and class of models if one desires to re-scale the approximate posterior distribution to calibrate the uncertainty intervals.

\section{Application of simulation-based calibration to real data}\label{sec:application}

\subsection{Model fit checking and calibration procedure}\label{sec:simdesc}
We now apply the calibration procedure to real data from the Current Employment Statistics (CES) survey, where we use our area-level mixture model described in Section \ref{sec:models} to obtain better estimates of employment levels in 166 domains within the ``Leisure" industrial sector. As a first step, we fit the model and extract draws from the model posterior predictive distribution of $y_i$ and $v_i$. 

Before performing simulation-based calibration, we conduct posterior predictive checks to assess the correctness of the model formulation based on how well it preserves the distribution properties of the real data.  We examine the marginal distributions over domains $i \in (1,\ldots,N$) for replicate draws of each of $(y^{(\alpha)}_{i})_{i=1}^{N}$ and $(v^{(\alpha)}_{i})_{i=1}^{N}$ as well as examine the joint distributions, $(y^{(\alpha)}_{i},v^{(\alpha)}_{i})_{i=1}^{N}$.  We only expect our simulation-based calibration procedure to work if the specified model well reproduces the distribution properties of the real data.

\begin{figure}[h!]
  \centering
  \begin{subfigure}{0.45\textwidth}
    \includegraphics[width=\textwidth]{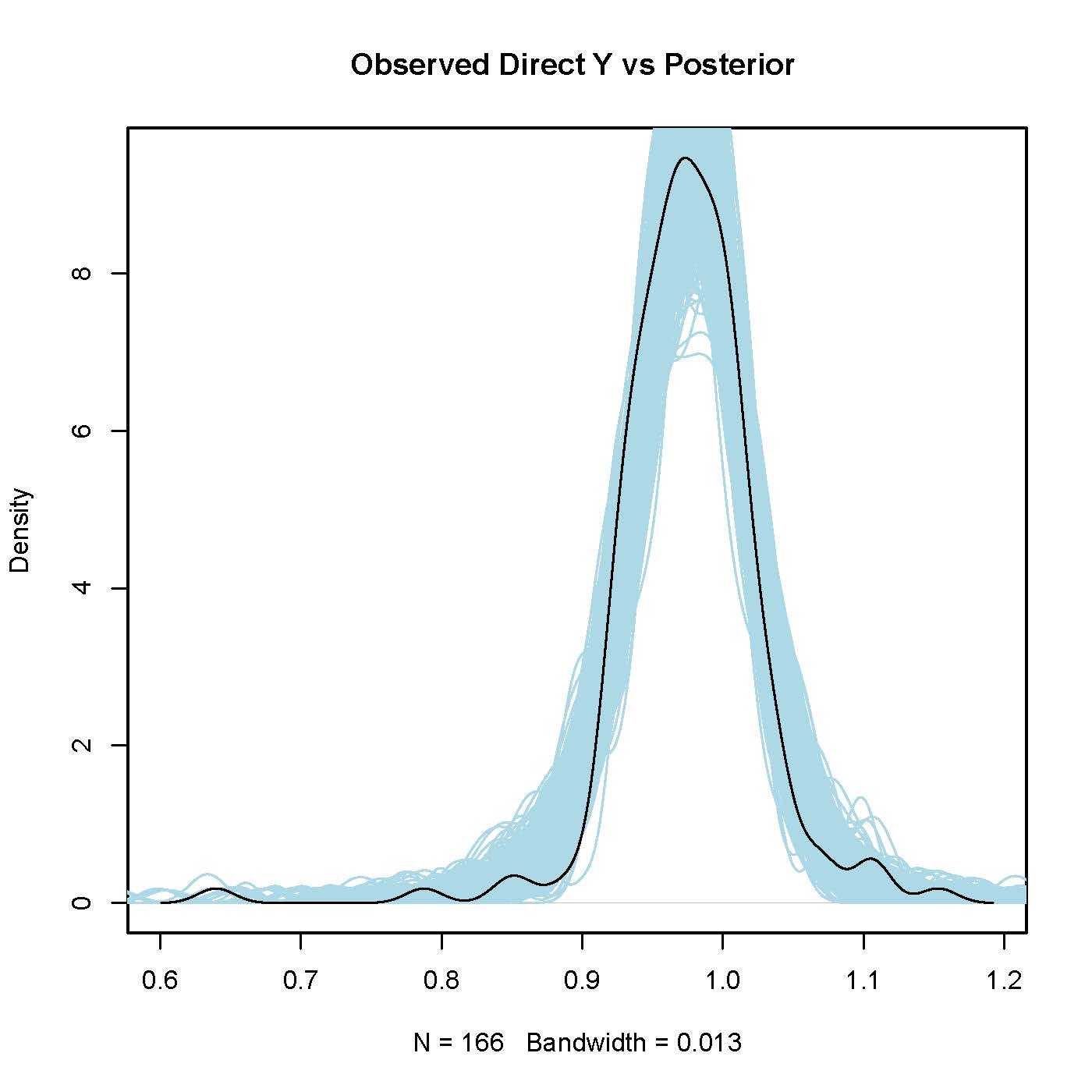}%
  \end{subfigure}
  \begin{subfigure}{0.45\textwidth}
    \includegraphics[width=\textwidth]{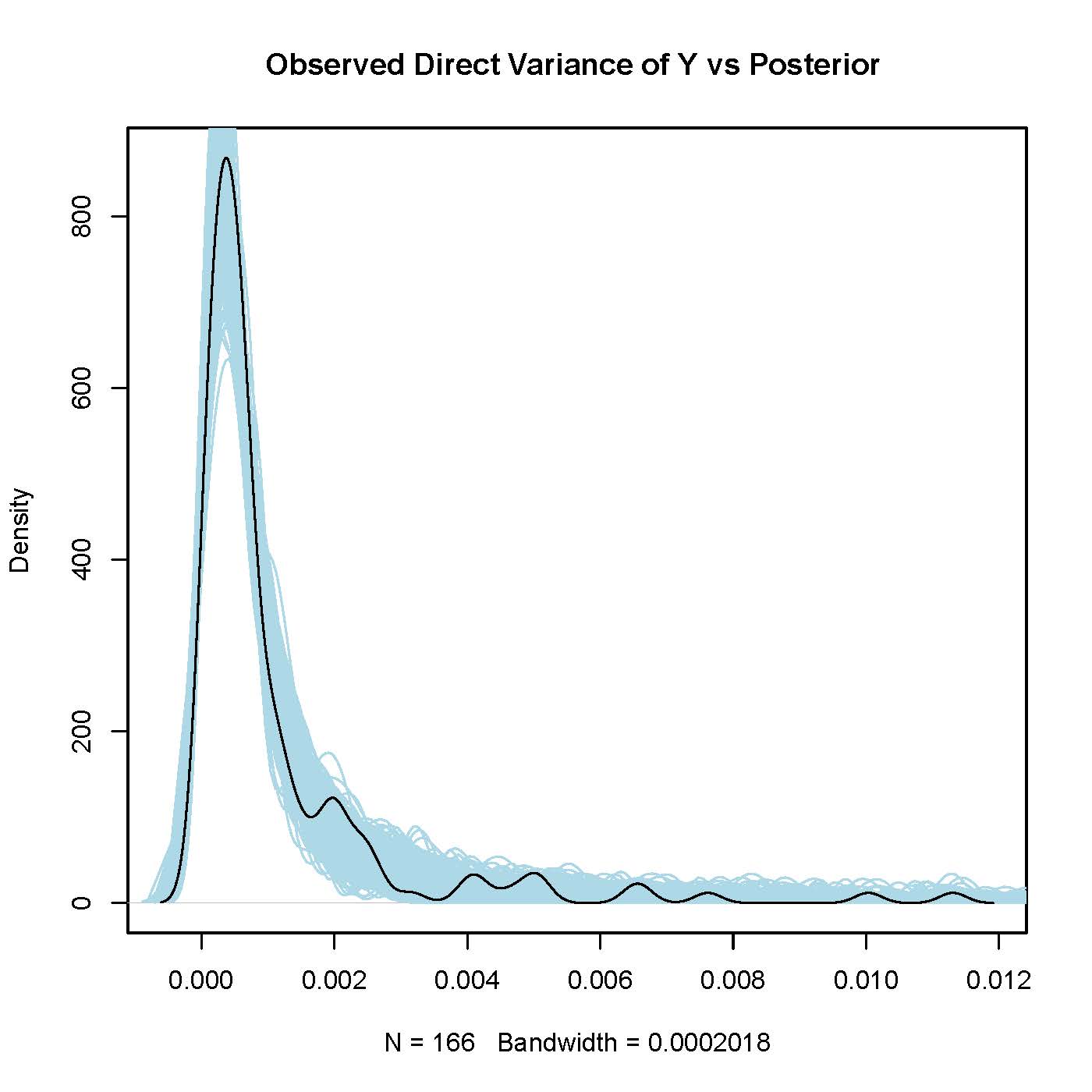}%
  \end{subfigure}
\caption{Posterior predictive check for (a) direct estimates $y_i$ and (b) observed variances of $v_i$ of direct estimates $y_i$.}
  \label{fig:ppc}
\end{figure}
\FloatBarrier

\begin{figure}[h!]
\centering
  \includegraphics[width=0.7\linewidth]{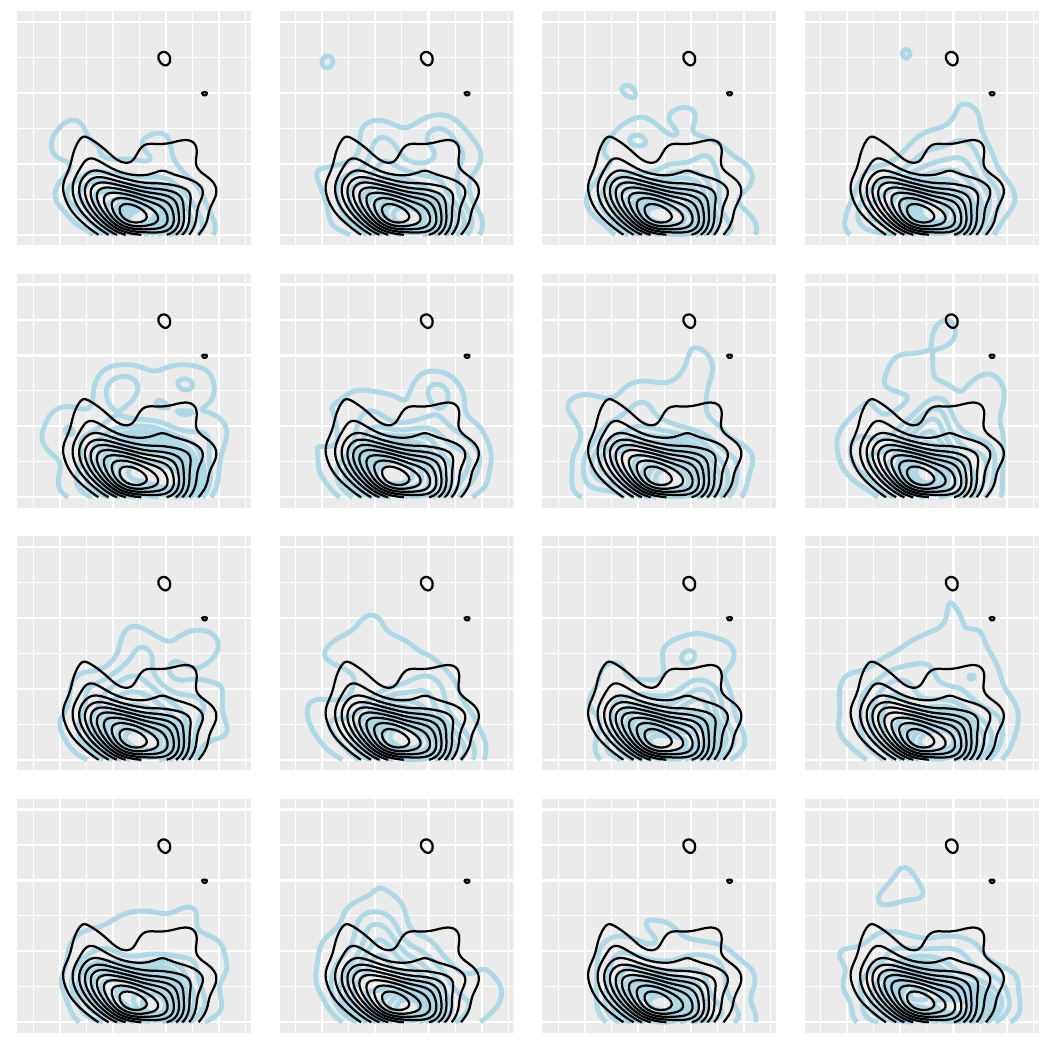}
  \caption{Posterior predictive check for pairs $(y_i,v_i)$, based on 16 random draws from the posterior predictive distribution. Each of the 16 panels represents a single vector draw (in light blue) from the posterior predictive distribution, overlayed by the observed distribution (black) over domains.}
  \label{fig:contour}
\end{figure}
\FloatBarrier

Plots in Figure \ref{fig:ppc} demonstrate that the model provides reasonable fit for observed $y_i$ and $v_i$.  Next, we briefly check if the model also preserves correlation between vectors $y_i$ and $v_i$. For this, we use a contour plot to show the density of two-dimensional vector $(y_i,v_i)$ over the set of domains. On the same plot, we include the analogous density of vector $(y_i^{(d)},v_i^{(d)})$  obtained by randomly drawing from the posterior predictive distribution of $(y_i,v_i)$. The plot panels in Figure \ref{fig:contour} show $D = 16$ sets of such contour plots corresponding to $D = 16$ draws from the posterior predictive distribution overlayed by the observed distribution of $(y_i,v_i)$. The contour plots demonstrate that correlation between $y_i$ and $v_i$ in the simulated data is similar to the observed data. We, therefore, conclude that the model describes the data reasonably well.  

We now save $A=100$ random draws $({y}_i^{(\alpha)},{v}_i^{(\alpha)})$, $\alpha = 1,\ldots,A$, from the posterior predictive distribution of $({y}_i,{v}_i)$ and compute calibration adjustments as described in Section \ref{sec:calibration}. 

There is a practical concern that it would not be feasible to perform calibration in the tight real-time monthly CES production environment. To overcome this difficulty, we conjecture that the CES calibration adjustments would be similar for each month in a given industry because the underlying model formulations and data structures are highly similar. If this is the case, our plan for the real-time application of the method would be to derive a set of adjustments once a year based on past year historical data, before the start of a new production cycle for the current year. These historical-based monthly adjustments would be averaged over the 12 months of the previous year, and the average values would be applied in the upcoming monthly production cycle. 

To test the proposed procedure, we save another $B=100$ sets of draws from our initial model run, in addition to the $A=100$ sets already used to derive the calibration adjustments. We then run the model once for each of the $B$ datasets and apply our average adjustments to check coverage properties. The procedure is summarized below.

\begin{enumerate}
    \item Use historical CES monthly data and fit the model for each month of the year. Call it the ``initial" model run. 
    \item Produce $A=100$ re-sampled datasets using draws from the initial model run. Derive calibration adjustments from these datasets, as described in Section \ref{sec:calibration}. Run the calibration procedure for each month of a year (to obtain calibrated intervals for each month). 
    \item Produce additional $B=100$ ``test" datasets using $B$ additional draws from the initial model run. Fit the model once on each of the $b=1,\ldots, B$ test datasets using the same algorithm as in the initial model fit. For each model run, $b$, we use the associated  $({y}_i^{(b)},{v}_i^{(b)})$ as "observed'' data.  
    \item Apply calibration adjustments to estimates obtained from each of the $B$ model runs.
    \item For each $b$, construct confidence intervals for model fitted posterior means $m({\theta}_i^{(b)})$. Compute the number of times (over $B$) that respective true values ${\theta}_i^{(b)}$ are covered by the interval.  Perform the coverage computation for the originally fitted and adjusted estimates.
\end{enumerate}

We tested the average adjustment factors on our $B=100$ ``test" datasets and evaluated for each of the 12 months. Results are presented in Section \ref{sec:results}.

\subsection{Results}\label{sec:results}

We use draws from the VB posterior distribution of $\theta_i$ to construct credibility intervals; the adjusted confidence intervals are constructed using the rescaling  and pivotal approaches of Section \ref{sec:calibration}. For each domain $i$, we used adjustments averaged over 12 months and we test results separately for each month.  

\begin{table}[h]
\centering
\caption{Coverage properties of model fitted $\theta_i$, 50\% nominal, over 166 domains and 100 simulation runs}
\label{tab:covfittable_adj}
\begin{tabular}{c|ccc}
Month & Orig Fitted & Rescaled & Pivot  \\
\hline
Jan &  0.577 (0.750) & 0.523 (0.668) & 0.503 (0.642)   \\
Feb &  0.572 (0.582) & 0.523 (0.520) & 0.500 (0.500)   \\
Mar &  0.589 (0.531) & 0.537 (0.474) & 0.516 (0.455)    \\
Apr &  0.569 (0.663) & 0.517 (0.591) & 0.497 (0.566)   \\
May &  0.597 (0.511) & 0.544 (0.456) & 0.522 (0.437)   \\
Jun &  0.580 (0.573) & 0.529 (0.511) & 0.509 (0.491)  \\
Jul &  0.590 (0.585) & 0.538 (0.521) & 0.516 (0.500)   \\
Aug &  0.576 (0.710) & 0.520 (0.633) & 0.501 (0.608)   \\
Sep &  0.590 (0.469) & 0.538 (0.419) & 0.516 (0.402)    \\
Oct &  0.568 (0.661) & 0.516 (0.589) & 0.498 (0.565)    \\
Nov &  0.590 (0.608) & 0.540 (0.542) & 0.518 (0.520)    \\
Dec & 0.604 (0.542) & 0.549 (0.483) & 0.529 (0.464)    
\end{tabular}
\end{table}

\begin{figure}[h!]
\centering
  \includegraphics[width=0.7\linewidth]{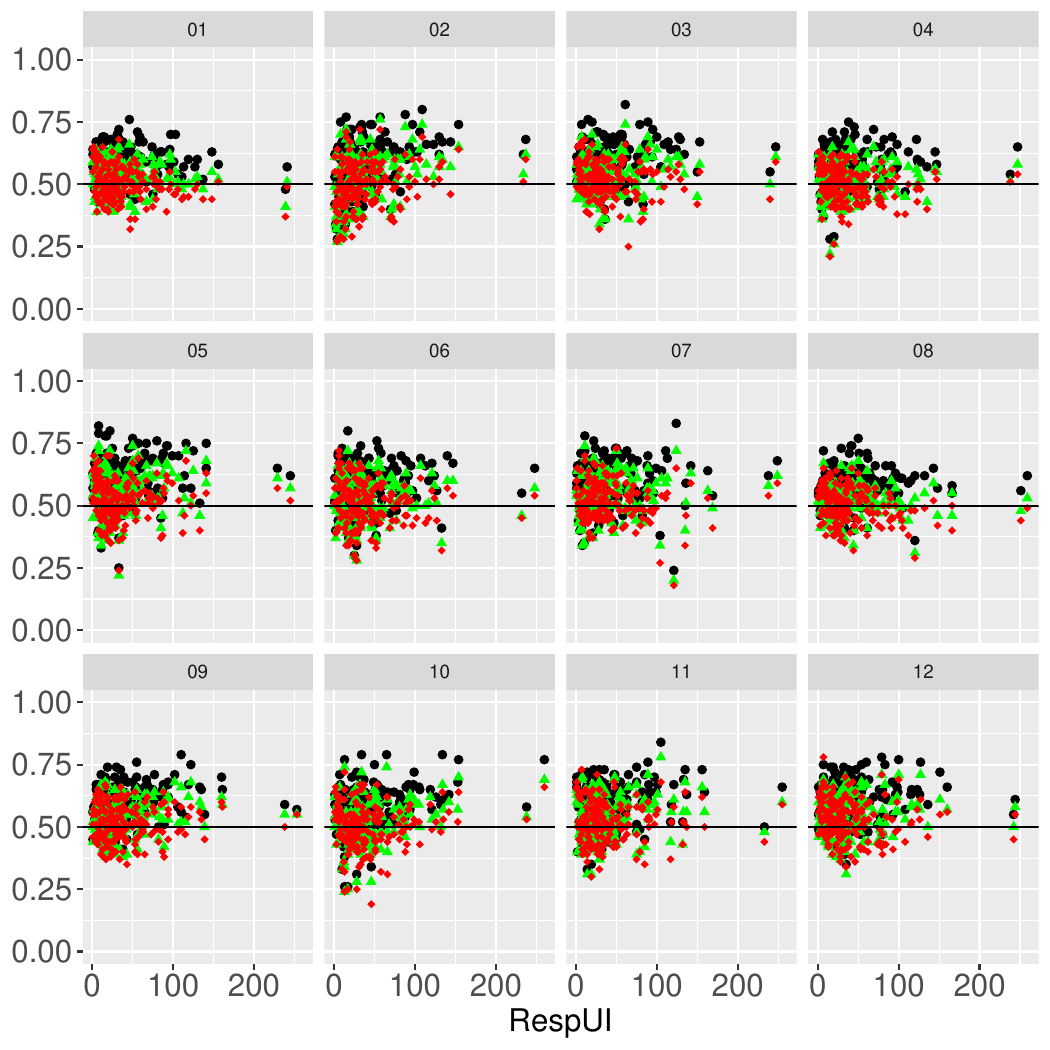}
   \caption{Domain coverages (50\% nominal) of fitted point estimates vs number of respondents for each of the 12 months. Black dots represent originally fitted model; green triangles represent adjusted coverages using rescaling method; red diamonds represent adjusted coverages based on the pivotal method. Domains in each panel are ordered along the horizontal axis by increasing number of sampled establishments from left-to-right. }
  \label{fig:coverage_mnth}
\end{figure}
\FloatBarrier

In Table \ref{tab:covfittable_adj}, we show monthly test results of coverages, averaged over domains. Respective average CI lengths are shown in parenthesis. We note the overcoverage of the original model CIs, ranging from $57$ to $60$ percent covered, for the nominal $50\%$. The rescaled CIs provide improved coverage (from $52$ to $55$ percent); the pivotal method gives best close to the nominal results and the shortest CI lengths. 

To complement the results averaged over domains  shown in Table \ref{tab:covfittable_adj}, we present monthly coverages for each domain in Figure \ref{fig:coverage_mnth}. Percent covered for each domain in a given month is plotted against the number of respondents. The black horizontal reference line corresponds to the $50\%$ nominal level. Black dots represent coverages corresponding to the original model fit. Similar to the simulations results of Section \ref{sec:simulations}, we observe overcoverage across the domains in the originally fitted model results. The rescaling method (green triangles) corrects the overcoverage to a certain degree. Red diamonds corresponding to the pivotal based adjustment method lie closer around the horizontal reference line, which is consistent with our conclusion that the pivotal based adjustment gives the best performing method among the alternatives considered.

\section{Discussion}\label{sec:discussion}
This paper formulates a parametric re-sampling procedure to calibrate the outputs of a scalable, approximate Bayesian estimation algorithm where the algorithm produces accurate point estimates (e.g., approximate posterior means) but \emph{incorrect} uncertainty quantification.  Our simulation-based calibration procedure produces asymptotically correct marginal posterior variances.   Our procedure requires that a scalable algorithm provide consistent point estimation (though there can be an asymptotically constant bias) under a Bayesian model that asymptotically contracts on the true generating model.   It would be expected that practitioners would only use a scalable algorithm to the extent that it produced similarly high quality point estimation as do Markov Chain Monte Carlo (MCMC) algorithms.

Bayesian hierarchical models are favored for small domain estimation in BLS for their ability to allow the data to estimate complex correlation structures among domains that regulates shrinking to produce more accurate domain point estimates.  A fast turnaround monthly BLS data publication cycle requires use of computationally scalable estimation algorithms.  BLS seeks to publish both smoothed domain point estimates, $\theta_{i}$, as well as their variances (of the $\theta_{i}$) so that users may assess the quality of each point estimate.   We use Stan's \citep{stan:2015} ADVI variational Bayes estimation algorithm and achieve high quality point estimation, but the variances are incorrect such that uncertainty intervals do not achieve nominal coverage.

Our application of the simulation-based calibration procedure requires re-estimation of the model on the order of $A = 100$ times, which reduces the scalability of the estimation algorithm (unless one possesses the computational resources to parallelize the $A$ model runs).  Yet, we show that under a regularly produced dataset for a collection of domains that computation of the mean and variance adjustment factors may be performed once and applied for each month throughout the year because the model formulation and data structures that together determine the calibration adjustments are largely unchanged over the months.  This procedure produces uncertainty intervals on a monthly basis that provide much better coverage than unadjusted intervals.

Our derivation of a calibration scale adjustment factor for each domain, $c_{i}$, by standardizing the distribution of a pivot statistic formed over replicate model runs provides a fast and easy numerical procedure.

Finally, an alternative simulation-based procedure calibrates the first and second moments using replicate data drawn from the \emph{prior} predictive distribution (as contrasted with our approach that calibrates data drawn from the approximate \emph{posterior} predictive distribution) \citep{talts2020validating}.   Calibrating from the prior predictive does not require an initial model estimation using the approximation algorithm as does our algorithm.  Yet, we find that while calibrating from the prior predictive achieves nominal coverage for simple models like the Fay-Herriot of Section~\ref{sec:FH}, it performs poorly (fails to achieve nominal coverage) for more complex models that co-estimate both the point estimates and variances as does our FHV and CFHV models of Sections~\ref{sec:FHS} and \ref{sec:models}.  We speculate that there is an implicit condition that the distribution of replicate data generated from the prior predictive distribution express some minimal mass covering the true data generating distribution, which is overly difficult to achieve in practice for a complex hierarchical model.  Even more, in the case of the FH where calibration from the prior predictive is feasible, it nevertheless produces longer interval lengths than calibrating from the posterior predictive where both algorithms achieve nominal coverage.

\acks{The authors would like to thank Andrew Gelman, Professor of Statistics in the Departments of Statistics and Political Science at Columbia University for suggesting the pursuit of a re-sampling based calibration procedure and for his coaching throughout the project development.}


\bibliography{refs}

\end{document}